# Half-Magnetic Topological Insulator

Ruie Lu[1*], Hongyi Sun[1*], Shiv Kumar[2*], Yuan Wang[1*], Mingqiang Gu[1], Meng Zeng[1], Yu-Jie Hao[1], Jiayu Li[1], Jifeng Shao[1], Xiao-Ming Ma[1], Zhanyang Hao[1], Ke Zhang[2], Wumiti Mansuer[2], Jiawei Mei[1], Yue Zhao[1], Cai Liu[1], Ke Deng[1], Wen Huang[1], Bing Shen[3], Kenya Shimada[2], Eike F. Schwier[2#], Chang Liu[1#], Qihang Liu[1,4#], Chaoyu Chen[1#]

[1] Shenzhen Institute for Quantum Science and Engineering (SIQSE) and Department of Physics, Southern University of Science and Technology (SUSTech), Shenzhen 518055, China.

[2] Hiroshima Synchrotron Radiation Centre, Hiroshima University, Higashi-Hiroshima, Hiroshima 739-0046, Japan.

[3] School of Physics, Sun Yat-Sen University, Guangzhou 510275, China.

[4] Guangdong Provincial Key Laboratory for Computational Science and Material Design, Southern University of Science and Technology, Shenzhen 518055, China.

*These authors contributed equally to this work.

#Correspondence should be addressed to E.F.S. (Eike.schwier@physik.uni-wuerzburg.de), C.L. (liuc@sustech.edu.cn), Q.L. (liuqh@sustech.edu.cn) and C.C. (chency@sustech.edu.cn)

**Abstract**

Topological magnets are a new family of quantum materials providing great potential to realize emergent phenomena, such as quantum anomalous Hall effect and axion-insulator state. Here we present our discovery that stoichiometric ferromagnet $MnBi_8Te_{13}$ with natural heterostructure $MnBi_2Te_4/(Bi_2Te_3)_3$ is an unprecedented "*half-magnetic topological insulator*", with the magnetization existing at the $MnBi_2Te_4$ surface but not at the opposite surface terminated by triple $Bi_2Te_3$ layers. Our angle-resolved photoemission spectroscopy measurements unveil a massive Dirac gap at the $MnBi_2Te_4$ surface, and gapless Dirac cone on the other side. Remarkably, the Dirac gap ($\sim$ 28 meV) at $MnBi_2Te_4$ surface decreases monotonically with increasing temperature and closes right at the Curie temperature, thereby representing the first smoking-gun spectroscopic evidence of magnetization-induced topological surface gap among all known magnetic topological materials. We further demonstrate theoretically that the half-magnetic topological insulator is desirable to realize the half-quantized surface anomalous Hall effect, which serves as a direct proof of the general concept of axion electrodynamics in condensed matter systems.

## I. Introduction

Magnetic topological insulators (TI) showcase quantum magnetism and nontrivial band topology, thereby providing a unique playground for exploring exotic quantum phenomena in condensed matter physics[1-16]. One paradigmatic example is the so-called axion insulator phase, which exhibits bulk topological magnetoelectric response with the phase angle $\theta = \pi$ protected by either inversion or time-reversal symmetry[10,15,16]. The resultant bulk-boundary correspondence is the predicted half-quantized surface Hall conductance $e^2/2h$ in the absence of external magnetic field given that a magnetic gap exists at the surface. The intrinsic magnetic TI $MnBi_2Te_4/(Bi_2Te_3)_n$ ($n = 1, 2, 3, ...$), comprising alternating layers of magnetic TI $MnBi_2Te_4$ and nonmagnetic TI $Bi_2Te_3$, holds the potential for realizing both the quantum anomalous Hall (QAH) insulator and the axion insulator phases[1-4,17-22]. In principle, this family of compounds are ideal candidates for the inversion-preserved axion insulators with a persistent surface gap originated from the long-range magnetic order. However, recent ARPES measurements unexpectedly reveal an almost gapless surface Dirac cone in $MnBi_2Te_4$[23-25]. Meanwhile, it remains controversial whether the topological surface states of two other members in $MnBi_2Te_4/(Bi_2Te_3)_n$ ($n = 1$ and 2) are gapped or gapless[26-34]. Unfortunately, none of these reported gaps have been proved to originate from magnetic orders since they remain open within the high-temperature paramagnetic (PM) phase[28,35-37], which obsoletes the realization of the axion insulator phase.

Here, we experimentally verified that stoichiometric $MnBi_8Te_{13}$ ($MnBi_2Te_4/(Bi_2Te_3)_n$ with $n = 3$), with intrinsic ferromagnetic (FM) ground state, is a "*half-magnetic topological insulator*", in which the surface magnetization exists at the $MnBi_2Te_4$ septuple-layer (SL) surface but at neither of the three $Bi_2Te_3$ quintuple-layer (QL) terminated surfaces. Our angle-resolved photoemission spectroscopy (ARPES) measurements have directly revealed at the SL termination (S-termination) a ∼ 28 meV surface gap below the Curie temperature of $T_C = 10.5$ K, which decreases monotonically with increasing temperature and closes right at $T_C$ to form a gapless Dirac cone, proving its magnetic nature. These results represent the first direct spectroscopic evidence of magnetization-induced topological surface gap among all known magnetic topological materials. In sharp contrast, a gapless Dirac cone with negligible FM proximity is observed on the opposite surface terminated by the triple $Bi_2Te_3$ QLs, analogous to the situation in nonmagnetic TI $Bi_2Te_3$. Utilizing density functional theory (DFT) calculations, we find half-QAH conductivity well localized at the SL termination, regardless of the cleavage of the other termination. Therefore, the half-magnetic topological insulator provides an ideal platform for observing the half-QAH effect at a single surface and the related axion physics.

## II. MnBi₈Te₁₃ single crystal with FM ground state

Single crystal $MnBi_8Te_{13}$ has a trigonal structure[38] with a space group of $R\bar{3}m$. The lattice of $MnBi_8Te_{13}$ consists of one $MnBi_2Te_4$ SL and three $Bi_2Te_3$ QLs stacking alternately along $c$ axis (Figure 1a). These SLs or QLs are coupled through weak van der Waals (vdW) forces. Cleaving the single crystal perpendicular to $c$ axis could yield four possible terminations, namely, the S-termination, Q-termination, QQ-termination and QQQ-termination. The crystallinity was examined by X-ray diffraction (XRD). As shown in Figure 1b, all of the diffraction peaks, particularly the low angle ones, can be well indexed by the (00$l$) reflections with lattice parameter $c = 132.6$ Å, in agreement with previous report[39].

The temperature-dependent anisotropic magnetic susceptibility (Figure 1c) shows Curie−Weiss behavior above $150\,\text{K}$ (inset) with the characteristic temperature $\theta_{CW} = 20\,\text{K}$ and $15\,\text{K}$ for $H//c$ and $H//ab$, respectively, through a fitting with $\chi(T) = \chi_0 + C/(T - \theta_{CW})$. Around $T_C = 10.5\,\text{K}$, an FM transition was revealed by magnetic susceptibility (Figure 1c) and resistivity measurements (Figure 1d). The frustration parameter ($\theta_{CW}/T_C$) for $H//c$ was calculated to be ~2, indicating a moderate magnetic frustration. For $H//c$, the observed larger bifurcation between zero field cooling (ZFC) and field cooling (FC) magnetization (Figure 1c) and magnetic hysteresis (Figure 1e) indicate an easy axis along $c$ axis, and an Ising type exchange interaction between adjacent Mn layers. The saturation moment $M_{sa} = 3.58\,\mu_B/\text{Mn}$ is close to that of $3.56\,\mu_B/Mn$ in MnBi$_2$Te$_4$[40] and $3.5\,\mu_B/\text{Mn}$ in MnBi$_4$Te$_7$[32]. The above magnetic properties suggest an FM order with out-of-plane magnetic moment configuration in MnBi$_8$Te$_{13}$, in contrast to the A-type AFM ground states found in other MnBi$_2$Te$_4$/(Bi$_2$Te$_3$)$_n$ compounds ($n = 0, 1, 2$)[41,42].

The field-dependent Hall resistivity ($\rho_{xy}(H)$) and magneto-resistivity ($MR = \frac{\rho_{xx}(H)}{\rho_{xx}(0)} - 1$) are shown in Figure 1f, 1g and Figure S1. The negative slope of $\rho_{xy}(H)$ in Figure S1f indicates electron-type carriers, and the obvious anomalous Hall effect is observed for $H//c$. In a ferromagnet, the Hall resistivity is described by the formula $\rho_{xy} = R_0 H + \rho_{xy}^A = R_0 H + R_s M$, where $R_0$ is the ordinary Hall coefficient, $\rho_{xy}^A$ is the anomalous Hall resistivity, $R_s$ is the anomalous Hall coefficient and $M$ is the magnetization. Above $T_C$ (20 K), $\rho_{xy}(H)$ exhibits the same slope at all $H$ (see Figure S1f), indicating a constant $R_0$ which allows us to subtract the ordinary Hall resistivity to obtain the anomalous part as shown in Figure 1f. $R_s$ scales well with the $M$–$H$ curve to the anomalous part of the Hall resistivity and is calculated to be $R_s = 1.76 \times 10^{-6}\,\text{m}^3/\text{C}$, two orders of magnitude larger than $R_0 = 1.15 \times 10^{-8}\,\text{m}^3/\text{C}$. Unlike the previous report[39], the $MR$ from both increasing and decreasing field measurements keeps a near-vanishing value ($< 0.1\%$) and exhibits sharp peaks without any overlap at the appearance of anomalous Hall plateau. This feature is reminiscent of the $MR$ behavior in Cr-doped (Bi,Sb)$_2$Te$_3$ films when approaching the quantum anomalous Hall region[11,12].

### *III. Gapped and gapless TSS Dirac cone in MnBi$_8$Te$_{13}$*

We employ a $\mu$-Laser-ARPES system[43] with a focused laser spot size of ~5 μm to measure the termination-sensitive band structure of MnBi$_8$Te$_{13}$. Figures S2 and S3 present the spectra at a high-symmetry direction as well as a set of constant energy contours for all four terminations. Here in Figure 2 we highlight the band structure of the S-termination and its opposite cleaving plane, the QQQ-termination. Shown are spectra taken at 7 K and 20 K, which correspond to FM and PM phases, respectively. In the FM phase, the S-termination shows an unambiguous energy gap of about 28 meV at the Dirac point (Figure 2c). This is in sharp contrast to other Mn-Bi-Te family members such as MnBi$_2$Te$_4$[23-25,33], MnBi$_4$Te$_7$[27,33,35] and MnBi$_6$Te$_{10}$[31,33], whose S-terminations consistently show no apparent gap-opening at the Dirac point below the magnetic transition[44]. Above $T_C$ in the PM phase, a gapless Dirac cone is observed (Figure 2d). Comparison between the FM and PM phases suggests that the origin of the surface gap for the S-termination is magnetism.

The gap opening is captured by an effective massive Dirac Hamiltonian $H_{surf}(k) = (\sigma_x k_y - \sigma_y k_x) + m_{eff}\sigma_z$, where the first two terms describe a Dirac cone, and the last the effective Zeeman field induced by the ferromagnetically-ordered Mn atoms. The gap size, $m_{eff} \sim 0.28\ meV$, is in qualitative agreement with our DFT prediction (Figure 2b). The detailed comparison between the DFT and the ARPES results is provided in Figure S2.

At the QQQ-termination, the gapless Dirac surface states appear to persist below $T_C$. At first sight, this seems to contradict with the broken of time-reversal symmetry. However, given the considerable spatial separation between the top $Bi_2Te_3$ QL and the magnetic $MnBi_2Te_4$ SL, it is reasonable to assume a negligibly small effective Zeeman field for the surface states. Such a conjecture is indeed supported by our DFT calculation, which also reproduces the gapless Dirac cone despite a magnetic ground state (Figure 2f). We note here that the DFT surface-only spectra agree with the ARPES spectra better than the DFT bulk spectra, likely due to the limited photoemission probing depth[45]. Similar gapless Dirac cones have been observed by ARPES at the FM phase for both Q- and QQ-terminations, which is owing to the hybridization between the TSS and the bulk bands that buries the Dirac point[31], shown in Figure S4. In all, $MnBi_8Te_{13}$ is a half-magnetic topological insulator. The time-reversal symmetry is broken at the S-termination where shows a temperature-dependent gap, while approximately preserved at the other surface where shows a gapless state.

*IV. The nature of the TSS Dirac gap at S-termination*

Having established a TSS Dirac point gap opening in the S-termination of FM $MnBi_8Te_{13}$, we now demonstrate that this gap is indeed opened due to the long-range FM order of the magnetic moments. Zoom-in ARPES $k - E$ map of the S-termination in the PM state (15 K) is shown in Figure 3a while that in the FM state (7 K) is shown in Figure 3d. The magnetic-order-induced spectral change is concentrated at the Dirac point as highlighted in Figure 3b and 3e. For the PM S-termination, a gapless "X"-shape Dirac cone can be clearly resolved with its Dirac point being indicated by the black arrow in Figure 3b. The corresponding EDC spectra taken across the Dirac point can be fitted with two Lorentzian peaks, with the dominating one (dark blue) centered at the Dirac point energy $E_D \approx -0.21$ eV (Figure 3c).

For the FM S-termination, as presented in Figure 3e, the upper and lower Dirac cones are separated in energy by a sizable gap, with the two cones clearly detaching from each other. Fitting the corresponding EDC yields three Lorentzian peaks. The two dark blue peaks, locating at $E_1 \approx -0.19$ eV and $E_2 \approx -0.22$ eV, correspond to the upper Dirac cone minimum and lower Dirac cone maximum, respectively. These two peaks originate from the splitting of the gapless Dirac point peak centered at $E_D \approx -0.21$ eV (Figure 3c), resulting in a Dirac point gap of $\Delta = E_1 - E_2 \approx 28$ meV (Figure 3f). The light blue peak corresponds to a weak shoulder found both in PM and FM EDCs, whose peak position remains at the same energy at different temperatures and potentially originates from disorder.

In Figure 3g, systematic Lorentzian fitting to the $\bar{\Gamma}$ EDCs at various temperatures below and above the bulk PM-FM transition (10.5 K) are presented. The constant energy mapping and dispersions corresponding to each temperature are shown in Figure S5 and S6, allowing us to unambiguously extract the dispersion at the Γ point. At the lowest temperature (6 K), similarly three Lorentzian peaks are needed to fit the EDC, of which two dark blue peaks ($E_1$ and $E_2$) correspond to the split Dirac cone. The Dirac cone gap size $\Delta = E_1 - E_2$ and its temperature evolution is plotted in Figure 3h. With increasing temperature, $E_1$ and $E_2$ move closer to each other ($\Delta$ decreases) and eventually merge into one Lorentzian peak at 11 K (gap closes), strongly suggesting a clear correlation between the size of this Dirac point gap and the FM exchange interaction. It is worth noting that, while we can also assume similarly two dark blue peaks ($E_1$ and $E_2$) for the EDCs measured at $T \geq 11$ K, the fitting iterations always result in vanishing or even negative area of peak $E_2$, and $\Delta = E_1 - E_2 \leq 3$ meV, which is negligible compared to the width of the Lorentzian peaks. In short, the gapless-ness of the Dirac cone at temperatures above 11 K is well established.

Assuming a linear relation between this exchange splitting and the magnetic moment, the gap should be well described by a power law curve[46] $\Delta \sim E_0 \cdot (1 - T/T_0)^{2\beta}$, where $E_0$ represents the saturated exchange splitting energy at $T = 0$ K. Fitting the $\Delta(T)$ curve with this power law function yields $T_0 = 11 \pm 1$ K and $\beta = 0.23 \pm 0.02$. The fitted onset temperature $T_0$ matches the susceptibility-derived Curie temperature well within the fitting error. The saturated exchange splitting energy is fitted as $E_0 = 33 \pm 1$ meV. We thus established an FM-induced Dirac point gap in the S-termination of MnBi$_8$Te$_{13}$. It is noteworthy that, although Dirac point gaps have been reported for other members of the Mn-Bi-Te family[28,35-37], these observations are still controversial[23-25,27,31,33]. In particular, all the reported gaps remain open above the magnetic transition temperature, contradicting the scenario of the restoration of time reversal symmetry. Consequently, our results that a TSS Dirac cone gap decreases monotonically with increasing temperature and closes right at $T_C$ forming a gapless Dirac cone represent the first smoking-gun evidence of TSSs gapped by the magnetic order among all known magnetic topological materials.

*V. Surface anomalous Hall conductance as a signature of axion insulator*
Till now we have demonstrated a magnetic gap at the S-termination and gapless feature at the QQQ-termination of MnBi$_8$Te$_{13}$, rendering the material a "half-magnetic topological insulator". To further identify its topological nature, we next theoretically analyze the surface anomalous Hall conductance (AHC) of this gapped surface and corresponding experimental signatures. Due to the inversion symmetry, the band structure of MnBi$_8$Te$_{13}$ may be characterized by a higher-order topological invariant, *i.e.* the $Z_4$ number (the symmetry indicator of inversion[47,48]). Our explicit computation shows that MnBi$_8$Te$_{13}$ has $Z_4 = 2$, in agreement with a previous study[39] (Figure S8 and Table S1). For an FM compound, while $Z_4 = 1$ or 3 implies a Weyl semimetal, $Z_4 = 2$ corresponds to an axion insulator or a 3D Chern insulator, with distinct surface AHC behaviors[13]. Therefore, we compute the surface AHC by integrating the local Chern numbers through surface-related layers for two-dimensional slabs of MnBi$_8$Te$_{13}$, expressed as

$$\sigma_{xy}(L) = \frac{e^2}{h} \cdot \frac{-4\pi}{A} \text{Im} \sum_{l=0}^{L} \frac{1}{N_k} \sum_k \sum_{vv'c} X_{vck} Y^\dagger_{v'ck} \rho_{vv'k}(l), \qquad (1)$$

where $X$ and $Y$ are the position operators along $x$ and $y$ directions, respectively, which are directly computed from the velocity operators through $X(Y)_{vck} = \frac{\langle \psi_{vk}|i\hbar v_{x(y)}|\psi_{ck}\rangle}{E_{ck}-E_{vk}}$. The indices $v$ and $c$ denote the valence and conduction bands, respectively. $\rho_{vv'}(l)$ is the projection matrix onto the corresponding layer $l$, which implies a summation over all atoms within a vdW layer. To uncover the locality of the surface AHC, we construct two slabs with different thickness. For the S-termination of a 16-vdW-layer slab, when $E_F$ lies in the gap of this surface, the layer-integrated AHC reaches $\sim \frac{e^2}{2h}$ at the second layer from the S-termination (Figure 4a). On the other hand, the metallic QQQ-surface does not have a well-defined surface Chern number because $E_F$ inevitably cuts the surface bands (Figure S9). In comparison, the slab with 17 vdW layers is symmetric and has a global band gap, leading to a well-defined integer-quantized Chern number for the whole slab, i.e. $C = 1$. Figure 4b clearly shows that both top and bottom surfaces contribute a half-quantized AHC, while the internal layers do not contribute to the global AHC. In the bulk there is oscillation around $\frac{e^2}{2h}$ with a period of the unit cell thickness (four vdW layers) starting from the fourth layer from the surface. Therefore, the half-quantized AHC of MnBi$_8$Te$_{13}$ is a local quantity at the S-termination, indicating an axion insulator phase.

Though direct experimental measurement of the half-quantized surface AHC is challenging due to various reasons, including sample quality, possible electrode scattering, actual size of the surface gap and the electron chemical potential in the sample, numerical validation in MnBi$_2$Te$_4$/(Bi$_2$Te$_3$)$_n$[13] and an experimental proposal based on non-local transport measurement have recently been provided[14], with a hexagonal six-contact-probing set-up shown in Figure 4c. In the case of MnBi$_8$Te$_{13}$, as discussed above, magnetism opens a gap in the S-termination surface. In the transport measurement the half-quantized AHC acts as a chiral state carried by the hinge of this surface. We computed the spectral functions at the hinges and the center of the top surface terminated by SL (Figure 4d), validating the existence of the chiral hinge state at the S-termination. It clearly shows that although the center of the S-termination (point 2 in Figure 4c) has a gap, nonvanishing chiral states exist near the boundary formed by the $x - y$ top surface and the $x - z$ side surface, i.e. points 1 and 3, with opposite chirality. Such chiral hinge states are embedded within the side surface states, causing imbalanced density of states at the two sides of the Dirac cone. In this sense, MnBi$_8$Te$_{13}$ is an excellent candidate for observing the signature of the long-sought axion insulator and topological magnetoelectric effect.

**ACKNOWLEDGEMENTS**

We thank Ni Ni, Haizhou Lu and Hu Xu for helpful discussions. This work is supported by the Shenzhen High-level Special Fund (No. G02206304, G02206404), the Guangdong Innovative and Entrepreneurial Research Team Program (Nos. 2019ZT08C044, 2017ZT07C062 and 2016ZT06D348), the National Natural Science Foundation of China (NSFC) (Nos. 11674149, 11874195, 11804144 and 11804402), National Key R&D Program of China under Grant No. 2019YFA0704901, the Technology and Innovation Commission of Shenzhen Municipality (No. JCYJ20150630145302240 and No. KYTDPT20181011104202253), the Highlight Project (No. PHYS-HL-2020-1) of the College of Science at SUSTech, Guangdong Provincial Key Laboratory for Computational Science and Material Design under Grant No. 2019B030301001 and Center for Computational Science and Engineering of Southern University of Science and Technology. The ARPES measurements were performed with the approval of the Proposal Assessing Committee of the Hiroshima Synchrotron Radiation Center (Proposal Numbers: 19BG044, 19BU002, 19BU005 and 19BU012).

# APPENDIX

## A. Materials and Methods

### Sample growth

MnBi$_8$Te$_{13}$ single crystals were grown by the conventional high-temperature solution method using Bi$_2$Te$_3$ as the flux. Mn (purity 99.98%), Bi (purity 99.999%) and Te (99.999%) blocks were placed in an alumina crucible with a molar ratio of Mn: Bi: Te =1: 12.3: 19.4. Then the alumina crucible was sealed in a quartz tube under the argon environment. The assembly was first heated up in a box furnace to 950 ℃, held for 10 hrs, then cooled down slowly to 574 ℃ over 120 hrs. After this heating procedure, the quartz tube was taken out quickly and then decanted into the centrifuge to remove the excess flux from the single crystals. Because the temperature window for the growth of MnBi$_8$Te$_{13}$ is very narrow and Bi$_2$Te$_3$ is an inevitable byproducts, we cut the single crystals into thin pieces and checked by single-crystal x-ray diffraction on both sides to select only pure phase of MnBi$_8$Te$_{13}$ single crystals.

### Transport and magnetic measurements

The structure of the crystals was characterized by x-ray diffraction with Cu $K\alpha$ radiation at room temperature using a Rigaku Miniex diffractometer. The diffraction pattern can be well indexed by the (00$l$) reflections with $\Delta_{2\theta} \sim 2°$ for adjacent peaks especially at lower angles ($< 20°$). Resistivity measurements were performed by a Quantum Design (QD) Physical Properties Measurement System (PPMS) with a standard six-probe method, using a drive current of 8 mA. The current flows in the $ab$ plane and the magnetic field is perpendicular to the current direction. Magnetic measurements were performed using the QD Magnetic Property Measurement System (MPMS) with the Vibrating Sample Mangetometer (VSM) mode. Temperature dependent magnetization results were collected with an external magnetic field of 100 Oe, both along and perpendicular to the [001] direction of the sample.

### ARPES measurement

μ-Laser-ARPES[43] measurements were performed at the Hiroshima Synchrotron Radiation Center (HSRC), Hiroshima, Japan with a VG Scienta R4000 electron analyzer and a photon energy of 6.36 eV. The energy and angular resolution were better than 3 meV and less than 0.05°, respectively. Samples were cleaved *in situ* along the (001) crystal plane under ultra-high vacuum conditions with pressure better than $5 \times 10^{-11}$ mbar and temperatures below 20 K.

### First-principles calculations

DFT calculations[49,50] were conducted by using the projector-augmented wave (PAW) pseudopotentials[51] and exchange-correlation was described by the Perdew-Burke-Ernzerhof (PBE) version of the GGA functional[52,53] as implemented in the Vienna *ab-initio* Simulation Package (VASP)[54]. Considering the transition-metal element Mn in MnBi$_8$Te$_{13}$, PBE + U functional with $U = 5$ eV was used for Mn 3$d$ orbitals for all the results in this work[55]. The $k$-mesh, energy cutoff and total energy tolerance for the self-consistent calculations were $5 \times 5 \times 5$, 500 eV, and $10^{-5}$ eV, respectively. The experimental lattice constants ($a_0 = 4.37$ Å, and $c_0 = 132.32$ Å) were taken, while the atomic positions were fully relaxed until the force on each atom is less than $10^{-2}$ eV/Å. Spin−orbit coupling was included in the calculations self-consistently. We constructed Wannier representations by projecting the Bloch states from the first-principles calculations of bulk materials onto Mn-$d$, Bi-$p$, and Te-$s$ orbitals. The topological surface states as well as the surface anomalous Hall conductivity were calculated in tight-binding models constructed by these Wannier representations, as implemented in the WannierTools package[56-59].

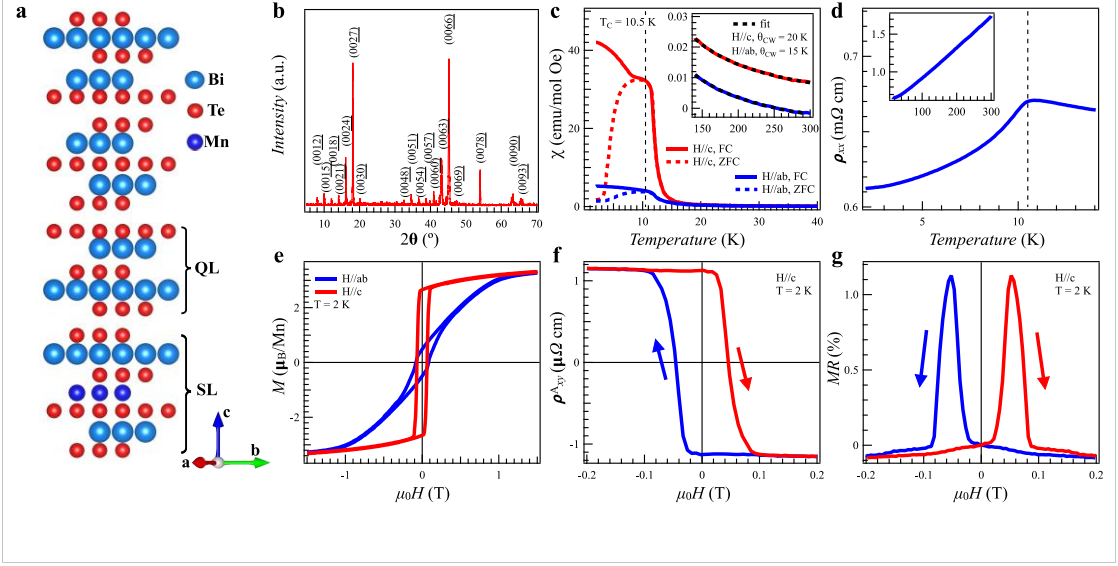

**Figure 1: Ferromagnetism and anomalous Hall effect in MnBi$_8$Te$_{13}$ single crystals.** (a), Schematic crystal structure with one unit of -SL-QL-QL-QL- sequences. (b), Single crystal x-ray diffraction result taken at 300 K. (c), Zero-field-cooled (ZFC, dashed) and field-cooled (FC, solid) magnetic susceptibility ($\chi$) vs temperature ($T$) for magnetic field $H = 100$ Oe parallel to the $ab$ plane (blue) and the $c$ axis (red), respectively. Inset shows the Curie−Weiss fitting for high temperatures (150 K − 300 K). (d), Zero-field in-plane resistivity ($\rho_{xx}$) vs $T$. Inset shows the results up to 300 K. (e), Field-dependent magnetization hysteresis at 2 K for $H//ab$ (blue) and $H//c$ (red). (f), Field-dependent anomalous Hall resistivity ($\rho^A_{xy}$) at 2 K for $H//c$. (g), Field-dependent transverse resistivity change ($MR = \frac{\rho_{xx}(H)}{\rho_{xx}(0)} - 1$) at 2 K for $H//c$.

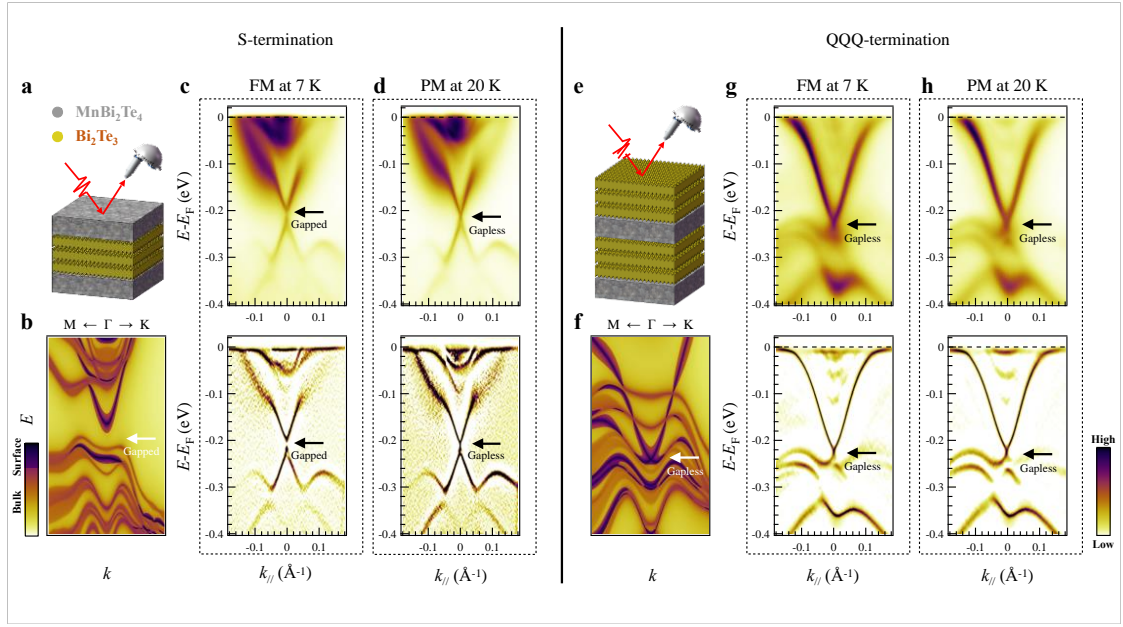

**Figure 2: Temperature evolution of the TSS Dirac cone gap at the S- and QQQ-termination of MnBi$_8$Te$_{13}$.** (a-d) ARPES results for the S-termination. (a), Schematic structural stacking configuration of the S-termination. (b), DFT band structure calculated for the FM state, with the surface-only states (dark violet to black) being superposed on the bulk states (yellow, orange and light violet). (c), Band structure along $\bar{M}-\bar{\Gamma}-\bar{M}$ measured at 7 K (FM state). (d), Band structure along $\bar{M}-\bar{\Gamma}-\bar{M}$ measured at 20 K (PM state). Data in (c) and (d) are shown in the form of the ARPES original spectra (top panel) and the 2D curvature spectra[60] (bottom panel). (e-h) Same as (a-d) but for the QQQ-termination. Clearly, entering the FM state opens a gap at the TSS Dirac cone for the S-termination, while no such gap is observed for the QQQ-termination.

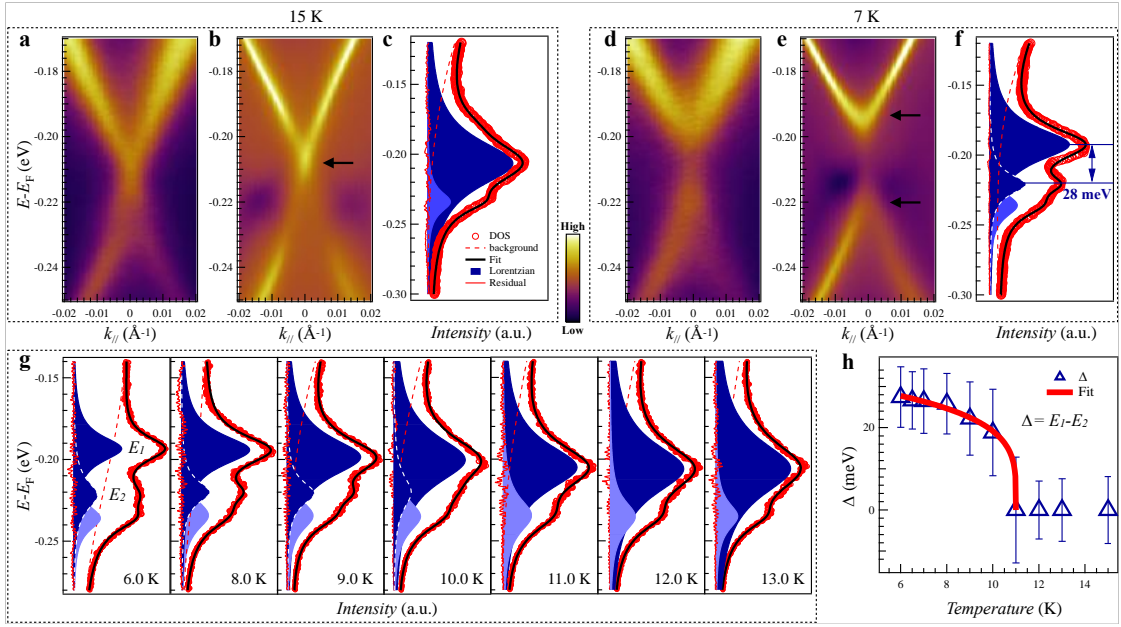

**Figure 3: Temperature dependence of the TSS gap in the S-termination of MnBi$_8$Te$_{13}$.** (a), Enlarged ARPES spectrum of the S-termination measured at 15 K (PM phase). (b), 2D curvature plot for the spectrum in (a). (c), Energy distribution curve (EDC) at $\bar{\Gamma}$ (integrated over a $\pm 0.001$ Å$^{-1}$ momentum window) and its fitting with multiple Lorentzian peaks. (d, e, f), Same plot as those in (a, b, c) but for data measured at 7 K (FM phase). (g), EDC fitting analysis at various temperatures, showing a clear gap opening for T ≤ 10.0 K. The EDCs for each temperature are extracted from the corresponding spectra shown in Figure S6 integrated over a $\pm 0.001$ Å$^{-1}$ momentum $k_{\parallel}$ window. Furthermore, the spectra for each temperature are extracted from the corresponding map shown in Figure S5 integrated over a $\pm 0.002$ Å$^{-1}$ momentum ($k_y$) window. (h), TSS Dirac cone gap size (blue triangles) evolution with temperature and its fitting (red solid line) using a power law curve. The error bar of the gap size is defined as $e = \sqrt{w_1^2 + w_2^2}$, where $w_1$ and $w_2$ represent the half-width-at-half-maximum for peak $E_1$ and $E_2$, respectively. We note that the EDC fitting in (g) yields standard deviation of the peak positions much smaller ($< 1\ meV$) than the error bars shown in (h).

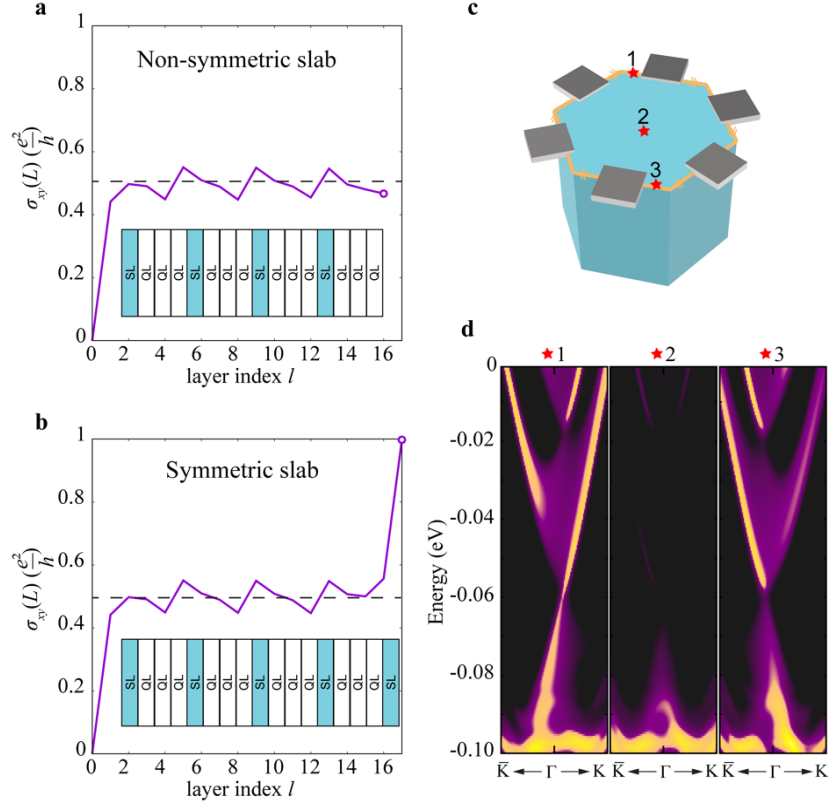

**Figure 4: Half-quantized surface AHC at the S-termination.** (a) and (b), Integrated layer-projected anomalous Hall conductance for the slabs with 16 and 17 vdW layers, respectively. The 16-layer slab is not symmetric, containing four unit cells as a half-magnetic topological insulator, while the 17-layer slab is symmetric with a global band gap and nontrivial Chern number, as shown in the insets. (c), A schematic plot for the non-local transport measurement with a hexagonal six-contact-probing set-up. (d), Spectral functions at three spots denoted in (c) showing chiral hinge states at the S-termination that manifest the half-QAH effect.